\documentclass{tMOP2e}
\usepackage{graphicx}

\newcommand{\bra}[1]{\langle #1|}
\newcommand{\ket}[1]{| #1 \rangle}
\newcommand{\braket}[2]{\langle #1|#2 \rangle}

\newcommand{\expi}[1]{ {\rm e}^{{\rm i} #1}}

\newcommand{\cre}[1]{\hat{#1}^\dagger}
\newcommand{\des}[1]{\hat{#1}}

\newcommand{\op}[1]{\hat{#1}}

\begin{document}

\doi{10.1080/0950034YYxxxxxxxx}
\issn{1362-3044}
\issnp{0950-0340}
\jvol{00} \jnum{00} \jyear{2008} \jmonth{10 January}

\markboth{Nunnenkamp and Rey}{Journal of Modern Optics}

\articletype{Snowbird 2008 -- Conference Proceedings}

\title{Macroscopic superposition states in rotating ring lattices}

\author{Andreas Nunnenkamp$^{b}$ $^{\ast}$\thanks{$^\ast$Corresponding author. Email: a.nunnenkamp1@physics.ox.ac.uk \vspace{6pt}} and Ana Maria Rey$^{a}$
\\\vspace{6pt}
$^{a}${\em{Institute for Theoretical Atomic, Molecular and Optical Physics, Cambridge, MA, 02138, USA}};
$^{b}${\em{Department of Physics, University of Oxford, Parks Road, Oxford, OX1 3PU, UK}}
}

\maketitle

\begin{abstract}
We investigate the effects of rotation on one-dimensional ultracold bosons confined to optical ring lattices.
First, we show that there exists a critical rotation frequency at which the ground state of a weakly-interacting and integer-filled atomic gas is fragmented into a macroscopic superposition state with different circulation.
Second, we point out several advantages of using slightly non-uniform ring lattices.
Finally, we demonstrate that different quasi-momentum states can be distinguished in time-of-flight absorption imaging and propose to probe correlations via the many-body oscillations induced by a sudden change in the rotation frequency.
\end{abstract}


\section{Introduction}

Macroscopic superposition states, to which we will refer as (Schr\"odinger) cat states, are one class of strongly-correlated states which has attracted much interest since the early days of quantum theory \cite{Schrodinger35}. Apart from foundational questions \cite{Leggett02} they have been shown to enable precision measurements at the Heisenberg limit \cite{Bollinger96} and serve as resources for various quantum information tasks \cite{Leibfried05}. Efficient cat state production is therefore an issue of theoretical and practical importance.

Among theoretical proposals for creating cat states with Bose-Einstein condensates \cite{Cirac98, Sorensen01, Dunningham06a, Dunningham06b} one is to load ultracold bosons into rotating ring lattices \cite{Hallwood06a, Hallwood06b, Rey07}.
In the first part of this paper, we investigate the effects of rotation on the ground state properties of bosonic atoms confined in a one-dimensional ring \cite{Hallwood06a, Rey07}. For commensurate filling and weak interactions there is a critical rotation frequency at which the ground and first excited states are the symmetric and anti-symmetric superposition states of opposite circulation. Their energy difference decreases exponentially with increasing number of particles \cite{Rey07}. This makes it hard to observe this effect even in systems with small numbers of particles.
On the other hand, superpositions of macroscopically distinct flow have been observed in experiments with superconducting quantum interference devices (SQUID) \cite{Friedman00, vanderWal00, Chiorescu03}.
This motivates the second part of this paper, where we point out that for a slightly non-uniform lattice the energy gap scales less severely with the number of particles \cite{Nunnenkamp08}.
Finally, we discuss how one can probe cat-like correlations via many-body oscillations induced by a sudden change in the rotation frequency and show that different quasi-momentum states can be distinguished in time-of-flight absorption images \cite{Nunnenkamp08}.

We note that there are related articles on the quantum dynamics of a vortex in a Bose gas
\cite{Watanabe2007} and topological winding in a 1D BEC \cite{Kanamoto2008}.

\section{Bosons in rotating ring lattices}

Optical ring lattices can be created experimentally by interfering a Laguerre-Gaussian (LG) laser beam with a plane wave co-propagating along the $z$-direction \cite{Amico05, Franke-Arnold2007}. By reflecting the combined beam back on itself, one can achieve confinement along the $z$-direction, thus creating an stacked array of disk shaped traps. By controlling the tunneling between the disks and making it much smaller than the corresponding tunneling within each ring, one can implement an array of effective 1D ring lattices. The lattices can be rotated by introducing a phase shift in one the two beams that compose the interference pattern.

We consider a system of $N$ ultracold bosons with mass $M$ confined in a 1D ring lattice of $L$ sites with lattice constant $d$. The ring is rotated in its plane (about the $z$-axis) with angular velocity $\Omega$. In the rotating frame of the ring the many-body Hamiltonian is given by \cite{Bhat2006, Rey07}
\begin{equation}
\hat {H}=\int d\mathbf{x}
\hat{\Phi}^{\dagger}\left[-\frac{\hbar^2}{2 M}\nabla^2 +
V(\mathbf{x}) +\frac{4 \pi \hbar^2 a}{2M}
\hat{\Phi}^{\dagger}\hat{\Phi}- \Omega \hat{L}_z\right]\hat{\Phi}
\end{equation}
where $a$ is the $s$-wave scattering length, $V(\mathbf{x})$ the lattice potential, $\hat{L}_z$ the angular momentum, and $\mathbf{x}$ the 3D spatial coordinate vector.

Assuming that the lattice is deep enough to restrict tunneling between nearest-neighbor sites and the band gap is larger than the rotational energy, the bosonic field operator $\hat{\Phi}$ can be expanded in Wannier orbitals confined to the first band $\hat{\Phi}=\sum_j \hat{a}_j W_j'(\mathbf{x})$ with $W_j'(\mathbf{x}) = \exp \left[ \frac{-i M}{\hbar} \int_{\mathbf{x}_j'}^\mathbf{x} \mathbf{A} (\mathbf{x}') \cdot d\mathbf{x}' \right] W_j(\mathbf{x})$. Here, $W_j(\mathbf{x})$ are the Wannier orbitals of the stationary lattice centered at the site $j$, $\mathbf{A}(\mathbf{x})=\Omega \hat{z} \times \mathbf{x}$ an effective vector potential and $\hat{a}_j$ the bosonic annihilation operator of a particle at site $j$. In terms of these quantities, the many-body Hamiltonian can be written, up to on-site diagonal terms which we neglect for simplicity, as \cite{Bhat2006, Hallwood06a, Rey07}
\begin{equation}
\op{H}
= -\sum_{j=1}^L \left( J_j \expi{\theta} \cre{a}_{j+1} \des{a}_j + H.c. \right) + \frac{U}{2} \sum_{j=1}^L \hat{n}_j(\hat{n}_j-1).
\label{ham}
\end{equation}
Here, $\hat{n}_j=\hat{a}_j^{\dagger}\hat{a}_{j}$ is the number operator at site $j$, $\theta$ is the effective phase twist induced by the gauge field, $\theta \equiv \int_{\mathbf{x}_{i}}^\mathbf{x_{i+1}} \mathbf{A}(\mathbf{x}')\cdot d\mathbf{x}' = \frac{M \Omega L d^2}{h}$, $J_j$ is the hopping energy between nearest-neighbour sites $j$ and $j+1$: $J_j \equiv \int d\mathbf{x} W_j^{*} \left[-\frac{\hbar^2}{2M} \nabla^2 + V(\mathbf{x})\right] W_{j+1}$, and $U$ the on-site interaction energy: $U\equiv \frac{4 \pi a \hbar^2}{M} \int d\mathbf{x}|W_j|^4$.

\section{The uniform ring lattice}

We start our discussion with the case of a uniform ring lattice, i.e.~$J_j = J$ for all $j$. To understand the effect of rotation on the ground state properties of atoms in a rotating ring lattice, it is convenient to write the many-body Hamiltonian (\ref{ham}) in terms of the quasi-momentum operators $\hat{b}_q = \frac{1}{\sqrt{L}} \sum_{j=1}^{L} \hat{a}_j e^{-2\pi i q j/L}$ where $2\pi q /(d L)$ is the discrete quasi-momentum and $q$ is an integer, $q=0,\dots, L-1$. In this basis the Hamiltonian (\ref{ham}) reads
\begin{equation}
\hat{H} = -2J \hat{H}_{\textrm{sp}}+\frac{U}{2L} \hat{H}_{\textrm{int}} =
\sum_{q=0}^{L-1} E_q\hat{b}_q^{\dagger}\hat{b}_{q}+\frac{U}{2L}
\sum_{q,s,l=0}^{L-1}\hat{b}_q^{\dagger} \hat{b}_s^{\dagger}
\hat{b}_l\hat{b}_{\|q+s-l\|_L}
\label{momehat}
\end{equation}
where $E_q=-2J \cos[2 \pi q/L-\theta] $ are single-particle energies and the notation $\|\quad \|_L$ indicates modulo $L$. The modulus is taken because in collision processes the quasi-momentum is conserved up to an integer multiple of the reciprocal lattice vector $2\pi/d$, i.e.~modulo Umklapp processes.

We begin by considering the non-interacting limit, $U=0$. While in a static ring, $\theta=0$, the ground state corresponds to all $N$ atoms in the $q_0=0$ quasi-momentum state, in a rotating ring the ground state acquires a finite quasi-momentum, $\frac{2\pi}{L} q_0$, to lower the energy gained by rotation. In other words it becomes a current carrying state. If one writes $\theta=\frac{2\pi}{L} m + \frac{ \Delta\theta}{L}$ with $m$ an integer and $0\leq \Delta\theta< 2\pi$, $q_0=m$ if $ 0\leq \Delta\theta< \pi$ and $q_0=m+1$ if $ \pi< \Delta\theta< 2\pi$. The quantum number $q_0$ can also be viewed as the winding number, which measures the circulation or "vorticity" of the gas: At the critical angular velocity $\Omega_c=\frac{\pi(2m+1) h}{M d^2 L^2}$, (i.e. $\Delta \theta=\pi$), the winding number increases by one unit.

In general, for $\Omega\neq \Omega_c$, the ground state is unique and corresponds to a macroscopically occupied state with well defined quasi-momentum: $q_0=m$ and $q_0=m+1$ for $\Omega < \Omega_c$ and $\Omega > \Omega_c$, respectively. However, the fact that exactly at the critical frequency, $\Omega=\Omega_c$, the single particle energies $E_{q=m}$ and $E_{q=m+1}$ become degenerate leads to a $(N+1)-$fold degenerate many-body ground state. The $N+1$ degenerate energy levels correspond to the states with $n$ atoms in $q=m$ and $N-n$ atoms in $q=m+1$, with $n=0,\dots,N$. We denote such states as $|n,N-n \rangle =\frac{1}{\sqrt{n!(N-n!)}} \hat{b}_m^{\dagger n}\hat{b}_{m+1}^{\dagger (N-n)}|0\rangle$.

In the presence of interactions, one expects the degeneracy to be lifted. For weak interactions, i.e.~$U/J \ll 1$, one can use lowest-order perturbation theory to account for their effect.
In view of the absence of direct coupling between the different $|n,N-n \rangle$ states, the energy shifts are diagonal and given by $E_{n}^{(1)}= N E_m+\frac{U}{2L}[(N(N-1)-2n(n-N)]$.
Therefore, in lowest order of perturbation theory, the states with minimal energy become those corresponding to $n=0$ and $n=N$. These two states remain degenerate and higher-order perturbation theory is needed to break the degeneracy. At this point one has to distinguish whether the number of atoms $N$ is commensurate (CO) or incommensurate (ICO) with the number of lattice sites $L$.

In the ICO case, there is no coupling between $|N,0\rangle$ and $|0,N\rangle$ and therefore, these states remain degenerate for all values of $U$. In the CO system, on the other hand, there are many different paths that couple these two states, and the number of paths increases exponentially with $N$ and $L$. The absence of coupling in the ICO case can be intuitively understood by considering the total quasi-momentum $K = \frac{2\pi}{L}\|\sum _{q=0}^{L-1}q \hat{b}_q^\dagger \hat{b}_q\|_L$. The many-body Hamiltonian (\ref{momehat}) exhibits a block diagonal form if the quasi-momentum Fock states are ordered according to $K$. In the CO case, $N=n L$ with $n$ an integer, the states $|N,0\rangle$ and $|0,N\rangle$ have total quasi-momentum $K=\frac{2\pi}{L}\|m n L|_L=0$ and $K=\frac{2\pi}{L} \|(m+1) n L|_L=0$, respectively, and thus both of them belong to the $K=0$ block. On the other hand, in the ICO case, $N=n L+ \Delta N$, and hence the two states $|N,0\rangle$ and $|0,N\rangle$, belong  to different blocks as $K= \|m n L+ m\Delta N|_L \neq \|(m+1) n L+ (m+1) \Delta N\|_L$. Thus these two states are not coupled by interaction.

In the CO case, we construct an effective $2 \times 2$ Hamiltonian by projecting on the subspace spanned by the two macroscopic states $|N,0\rangle$ and $|0,N\rangle$
\begin{equation}
\op{H}_{2 \times 2} =
\left( \begin{array}{cc}
E_{0}^{(1)}& \Delta \\
\Delta &  E_{N}^{(1)} \\
\end{array} \right)
\text{ with }
\Delta = A \left( \frac{U}{2L} \right)^{\bar{n}(L-1)}
\label{gap}
\end{equation}
where
$A=\sum_{i,j,\dots p} \frac{H_{0i} H_{ij}\dots H_{pN}}{(E_{0}^{(1)} -\varepsilon_i) (E_{0}^{(1)} -\varepsilon_i) (E_{0}^{(1)}-\varepsilon_j) \dots ( E_0^{(1)} - \varepsilon_p)}$
and $\bar{n}=N/L$ is the number density. The $H_{ij}$ are transition matrix elements introduced by the interaction term of the Hamiltonian $\op{H}_{\textrm{int}}$; and $\varepsilon_i$ are either $E^{(1)}_n$ or non-interacting many-body eigenenergies depending upon whether the intermediate states are or are not in the $|n,N-n\rangle$ manifold. The factor ${\bar{n}(L-1)}$ corresponds to the minimum number of collision processes necessary to connect the states $|N,0\rangle$ and $|0,N\rangle$ and the sum is over all the different paths that generate such couplings. At the critical rotation frequency we have $E_{0}^{(1)} = E_{N}^{(1)}$ and due to the non-zero value of $\Delta$ the symmetric and anti-symmetric superpositions $| \pm \rangle$ become the ground and first excited state separated from each other by an energy gap $2 \Delta$
\begin{equation}
| \pm \rangle = \frac{|0,N\rangle \pm |N,0\rangle}{\sqrt{2}}.
\label{cat}
\end{equation}

\section{Extension to a ring superlattice}

In the previous section we found that for a uniform ring lattice the coupling $\Delta$ between the two macroscopic states scales exponentially with increasing number of particles $N$ (see Eq.~(\ref{gap})), and that the cat state (\ref{cat}) only appears in the CO case. In this section we explore whether the situation can be improved by introducing a lattice modulation, i.e.~we generalize our discussion to ring superlattices where $J_j = J$ for $j$ even and $J_j = t$ for $j$ odd.

Due to the superlattice potential the quasi-momentum states $\ket{q}$ and $\ket{q+L/2}$ are coupled and the single-particle Hamiltonian is no longer diagonal in the quasi-momentum basis.
Via a unitary basis transformation
\begin{equation}
\left( \begin{array}{c} \op{c}_q \\
\op{c}_{q+L/2} \end{array} \right)
= \left( \begin{array}{cc}
\cos \alpha_q & i \sin \alpha_q \\
i \sin \alpha_q & \cos \alpha_q
\end{array} \right)
\left( \begin{array}{c} \op{b}_q \\
\op{b}_{q+L/2} \end{array} \right)
\label{definec}
\end{equation}
with
\begin{equation}
\tan 2\alpha_q = \frac{J-t}{J+t} \cdot \tan\left(\theta - \frac{2\pi q}{L}\right)
\end{equation}
we diagonalize the single-particle Hamiltonian $\op{H}_{\textrm{sp}}
= \sum_{q=0}^{L/2-1} \left( E_{q}^- \des{n}_q + E_{q}^+ \des{n}_{q+L/2} \right)$
where the single-particle energies are given by
\begin{equation}
E_{q}^{\pm} = \pm \sqrt{J^2 + t^2 + 2J t \cos \left(2\theta-\frac{4\pi q}{L}\right)}.
\label{spspectrum}
\end{equation}
The single-particle spectrum (\ref{spspectrum}) contains a degeneracy at the critical phase twist $\theta = \pi/4$ even the presence of the superlattice potential, i.e.~we have $E_0^- = E_{L/4}^-$, but note that for $L \not= 4$ these states are not the ground states of the system.

Following similar arguments as in the previous section we will show that weak on-site interactions lift the degeneracy at $\theta=\pi/4$ and lead to the formation of strongly-correlated states in the many-body system. In the absence of interactions $U = 0$ the ground state of a bosonic many-body system is the state with all bosons occupying the lowest-energy single-particle state. At the critical phase twist $\theta = \pi/4$ the single-particle spectrum has, however, a twofold degeneracy, so that there is a $N+1$-dimensional degenerate subspace at $N$-particle level. A convenient basis for this subspace is given by the Fock states $|n,N-n\rangle$ with $0 \le n \le N$, where $n$ particles are in the single-particle state of energy $E_0^{-}$ and $N-n$ particles in the one of energy $E_{L/4}^{-}$, respectively. For weak interactions $NU/L \ll 2\sqrt{J^2+t^2}$ this subspace is the low-energy sector of the many-body problem for all phase twists $\theta$ and tunneling strength ratios $t/J$.

The effective Hamiltonian in first order of the on-site interaction strength $U$ is
\begin{equation}
\op{H}_{\textrm{eff}} = \left(E_{0}^- \op{n}_0 + E_{L/4}^- \op{n}_{L/4} \right)
+ \frac{U}{2L} \left( 2 \op{n}_0 \op{n}_{L/4} + N^2 - N \right)
+ \left( \frac{i \eta U}{2L} \cre{c}_0 \cre{c}_0 \des{c}_{L/4} \des{c}_{L/4} + H.c. \right)
\label{heff}
\end{equation}
where $\op{n}_q = \cre{c}_q \des{c}_q$ are the number operators and the parameter $\eta$ simplifies for $\theta=\pi/4$ to $\eta = (J^2-t^2)/(J^2+t^2)$. At the critical phase twist $\theta = \pi/4$ the terms in the first bracket are an unimportant zero-energy offset, whereas the terms in the second bracket shift the energies of the states in the subspace differently, e.g.~they lead to an energy difference of $U(N-1)/L$ between the states $\ket{N,0}$ and $\ket{N-1,1}$, while the states $\ket{n,N-n}$ and $\ket{N-n,n}$ remain pairwise degenerate. The terms in the third bracket are off-diagonal in the Fock basis of the subspace and describe two-particle scattering between the two single-particle modes. They lift the remaining pairwise degeneracies in the many-body spectrum.

Let us now focus on slightly non-uniform rings $t/J \approx 1$ close to the critical phase twist $\theta \approx \pi/4$. Since the terms in the second bracket of Eq.~(\ref{heff}) increase the energy for all states in the subspace apart from $\ket{N,0}$ and $\ket{0,N}$ and the coupling between the states is weak (as the coupling $\eta$ is small in this limit), we project the effective Hamiltonian (\ref{heff}) onto the subspace spanned by these two nearly-degenerate lowest-energy states \cite{Rey07}. As there is no direct coupling between $\ket{N,0}$ and $\ket{0,N}$ we calculate the total coupling through intermediate states using perturbation theory. After eliminating the intermediate states we obtain, up to a constant energy shift which we neglect, the following $2 \times 2$ Hamiltonian
\begin{equation}
\op{H}_{2 \times 2} = \left( \begin{array}{cc}
\Delta E/2 & \Delta \\
\Delta^* & - \Delta E/2
\end{array}
\right)
\label{ham2by2}
\end{equation}
where $\Delta E$ is the energy difference between the states $\ket{N,0}$ and $\ket{0,N}$ caused by the detuning of the phase twist from resonance $\Delta \theta = \theta - \pi/4$, i.e.
\begin{equation}
\Delta E = N(E_{L/4}^- - E_0^-) \approx \frac{4J t N \Delta \theta}{\sqrt{J^2 + t^2}},
\end{equation}
and $\Delta$ is the coupling between the states $\ket{N,0}$ and $\ket{0,N}$ due to the off-diagonal terms of the effective Hamiltonian (\ref{heff}). As the latter only directly couples the states $|n,N-n\rangle$ and $|n\pm2,N-n\mp 2\rangle$, the first non-vanishing order is given by
\begin{equation}
\Delta = \frac{\bra{N,0} \op{H}_{\textrm{eff}}^{N/2} \ket{0,N}} {\prod_{j=1}^{N/2-1}(E_0^{(1)}-E_{2j}^{(1)})} = \frac{U}{L} \cdot \left(\frac{i \eta}{2}\right)^{N/2} \cdot \frac{N!}{\prod_{j=1}^{N/2-1} (2j)^2}
\label{gap2}
\end{equation}
where $E_j^{(1)} = \frac{U}{2L} \left( 2 j (N-j) + N^2 - N\right)$ is the diagonal interaction energy shift.

\begin{figure}
 \centering
 \textbf{(a)} \includegraphics[width=0.4\columnwidth]{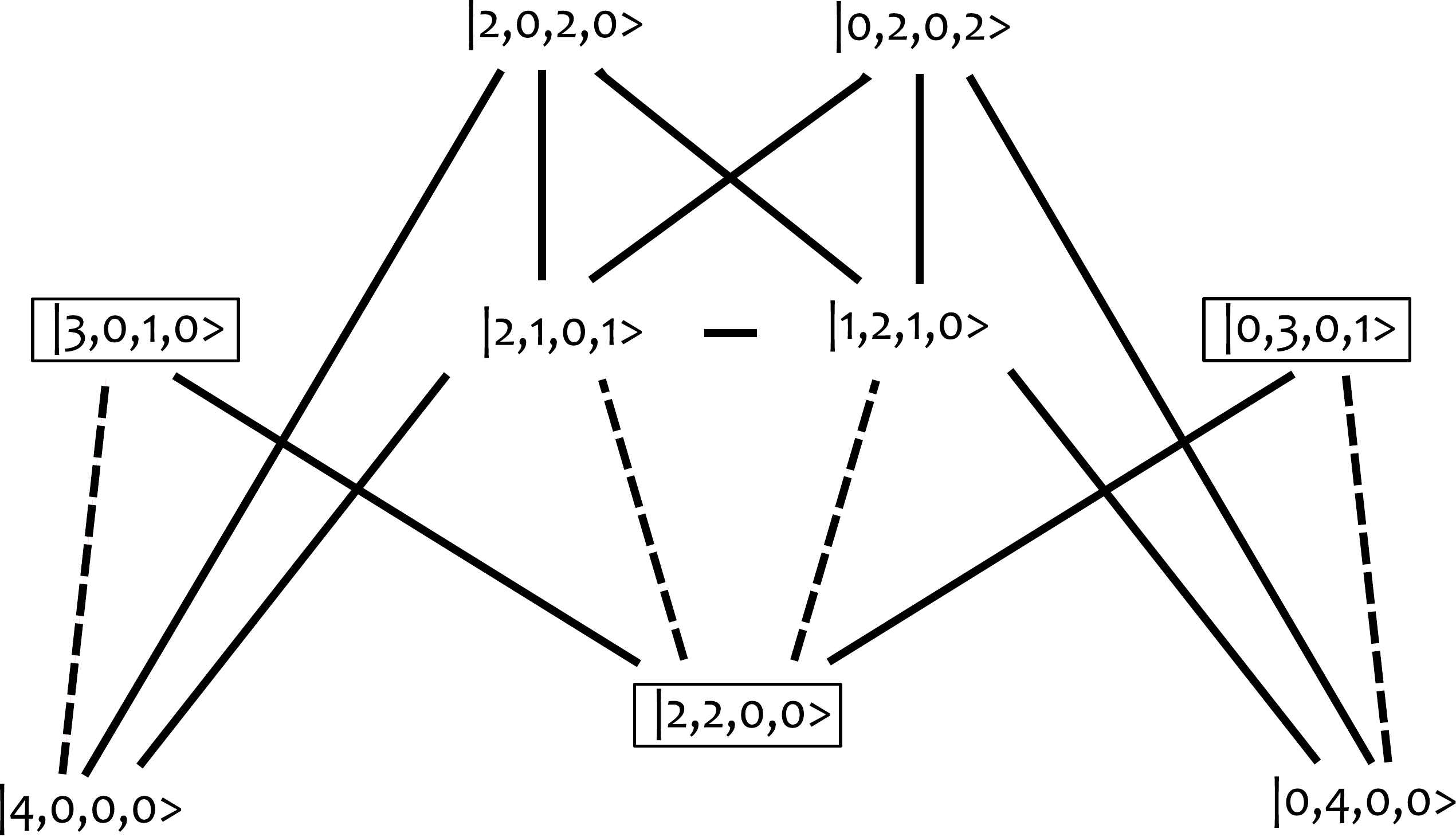}
 \hfill
 \textbf{(b)} \includegraphics[width=0.4\columnwidth]{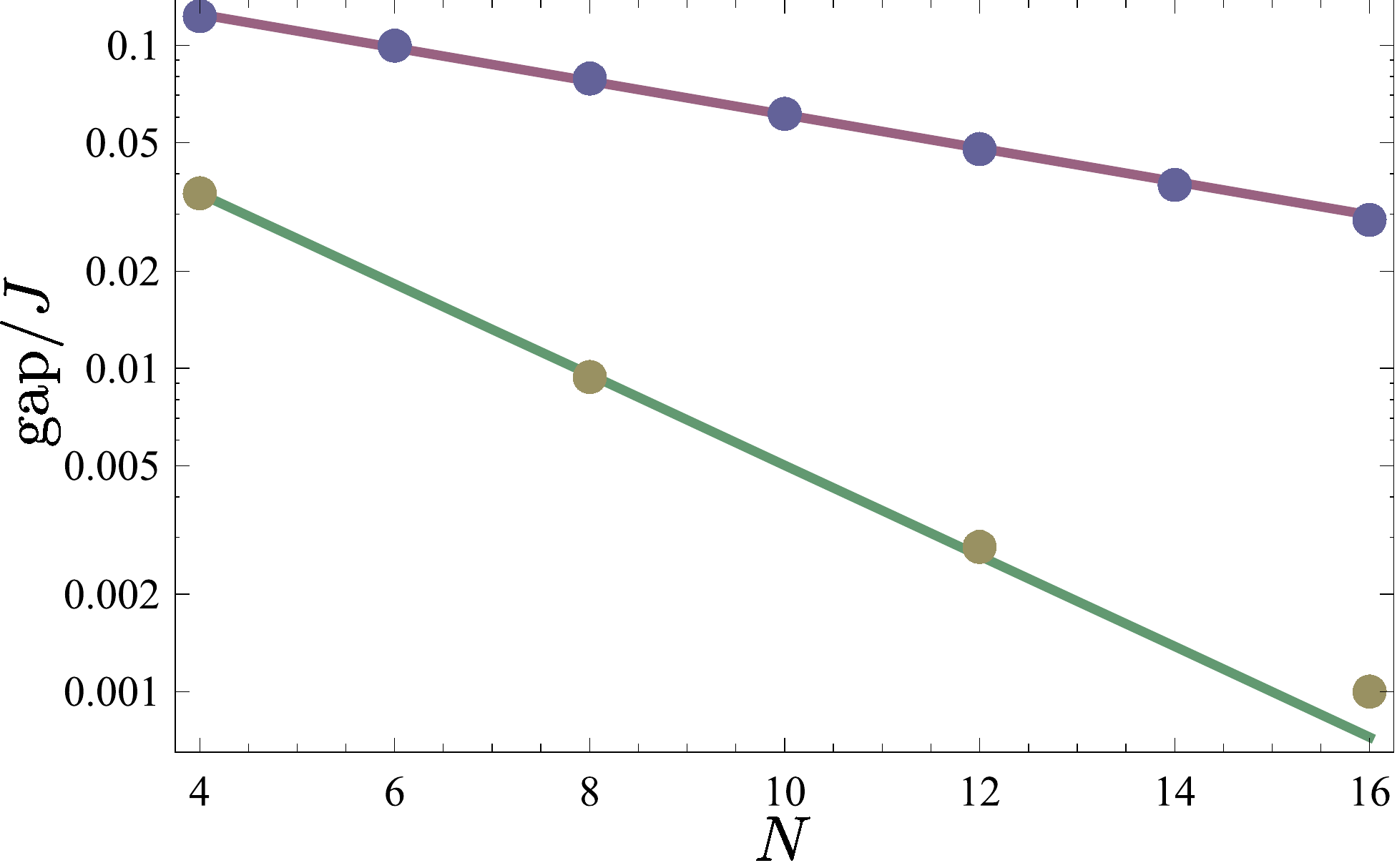}
 \caption{
 \textbf{(a)} Non-zero matrix elements between the quasi-momentum states $\ket{n_{q=0}, n_{q=1}, n_{q=2}, n_{q=3}}$ for $N=4$ particles on $L=4$ lattice sites. Solid lines signify coupling due to scattering $\op{H}_\textrm{int}$ and dashed lines due to the off-diagonal part of the single-particle Hamiltonian $\op{H}_\textrm{sp}$ in case of the superlattice $(J\not= t)$. We only show states which contribute in lowest non-vanishing order of perturbation theory to the cat state at the critical phase twist. All states with(out) box have total quasi-momentum $K=2$ ($K=0$).
 \textbf{(b)} Energy gap as a function of particle number $N$ with $t/J = 0.7$ (upper curve) and $t/J =1$ (lower curve) for the ring of $L=4$ sites and $U/J =0.5$. The solid lines are exponential fits to guide the eye.}
 \label{fig:gapscaling}
\end{figure}

In order to understand the influence of the superlattice potential on the formation of the cat state, we show in Fig.~\ref{fig:gapscaling}(a), for the example of $N=4$ particles on $L=4$ lattice sites, the \textit{quasi-momentum} states $\ket{n_{q=0}, n_{q=1}, n_{q=2}, n_{q=3}}$ which contribute to the ground and first excited state in lowest non-vanishing order of perturbation theory. Solid lines signify coupling due to the interaction Hamiltonian $\op{H}_\textrm{int}$ which is present in uniform ring lattices ($J = t$) as well as ring superlattices ($J \not= t$); dashed lines stand for the coupling due to the off-diagonal part of the single-particle Hamiltonian $\op{H}_\textrm{sp}$ induced by the superlattice. The three states $\ket{3,0,1,0}$, $\ket{0,3,0,1}$ and $\ket{2,2,0,0}$ have total quasi-momentum $K=2$ and are not coupled in the case of the uniform ring lattice. They introduce four efficient coupling paths involving only $N/2 = 2$ scattering processes and the intermediate state $\ket{2,2,0,0}$ lies in the degenerate subspace.

Let us compare our results for the coupling $\Delta$ in the case of a uniform ring lattice (\ref{gap}) with the case of a ring superlattice (\ref{gap2}). In both cases the coupling $\Delta$ decreases exponentially with the number of particles $N$. This is because multiple two-particle scattering processes are the microscopic origin of the coupling $\Delta$ and transitions between the two configurations $\ket{N,0}$ and $\ket{0,N}$ are thus highly off-resonant $N/2$-th and $\bar{n}(L-1)$-th order processes, respectively. Consequently, the energy gap vanishes in the thermodynamic limit $N \rightarrow \infty$ and cat state production both in uniform and slightly non-uniform ring lattices is restricted to modest numbers of atoms. However, as the energy gap in first non-vanishing order is proportional to $U(\eta/2)^{N/2}$ for slightly non-uniform rings as compared to $U(U/J)^{\bar{n}(L-1)}$ for uniform rings, it can be one order of magnitude bigger in the relevant parameter regime. Moreover, the appearance of cat states in slightly non-uniform rings is not limited to commensurate filling as long as the number of particles $N$ is even.

In Fig.~\ref{fig:gapscaling}(b) we plot the energy gap $L=4$ and $U/J = 0.5$ as a function of number of particles $N$ for $t/J=0.7$ and $t/J=1$. The solid line is an exponential fit to the points to guide the eye. We see that by changing from $t/J = 1$ to $t/J = 0.7$ the energy gap increases by over one order of magnitude for numbers of particles $N \ge 8$. As experiments are typically limited to $1 \, \textrm{sec}$ and typical hopping energies are of order of $0.05 E_r\sim 10^{2} \mathrm{Hz}$, the detection of cat states by non-equilibrium dynamics is limited to $\Delta/J =10^{-2}$. This in turn restricts our cat-state production scheme to a few tens of particles in contrast to less than ten particles for uniform ring lattices (see Fig.~\ref{fig:gapscaling}(b)).

\section{Dynamical detection of cat-like correlations}\label{secdy}

In this section we propose to induce many-body oscillations by suddenly changing the applied phase twist $\theta$ in order to detect the coherent superposition of two quasi-momentum states $\ket{N,0} = ( \cre{b}_0 )^N \ket{\textrm{vac}} / \sqrt{N!}$ and $\ket{0,N} = ( \cre{b}_{L/4} )^N \ket{\textrm{vac}} / \sqrt{N!}$ at the anti-crossing of the many-body spectrum.

If we assume the system is initially in the ground state of the full Hamiltonian at $\theta = 0$ which will predominantly be the state $\ket{N,0}$, i.e.~$\ket{\psi(t=0)} \approx \ket{N,0}$. Then, the phase twist is changed to $\theta = \pi/4$, where the eigenstates are approximately $\ket{\pm} \approx (\ket{N,0} \pm \ket{0,N})/\sqrt{2}$. The time evolution of the probability to be in the state $\ket{N,0}$ and $\ket{0,N}$ is
\begin{equation}
|b_0(t)|^2 = \left|\braket{N,0}{\psi(t)}\right|^2 \approx \frac{1 + \cos \nu t}{2}
\end{equation}
and $|b_N(t)|^2 \approx 1 - |b_0(t)|^2$ where $\hbar\nu = \sqrt{\Delta E^2 + |2\Delta|^2}$ denotes the energy gap and where the approximate signs indicate that we neglect the depletion due to the non-diagonal part of the single-particle Hamiltonian as well as the on-site interaction Hamiltonian.

\begin{figure}
 \centering
 \textbf{(a)} \includegraphics[width=0.4\columnwidth]{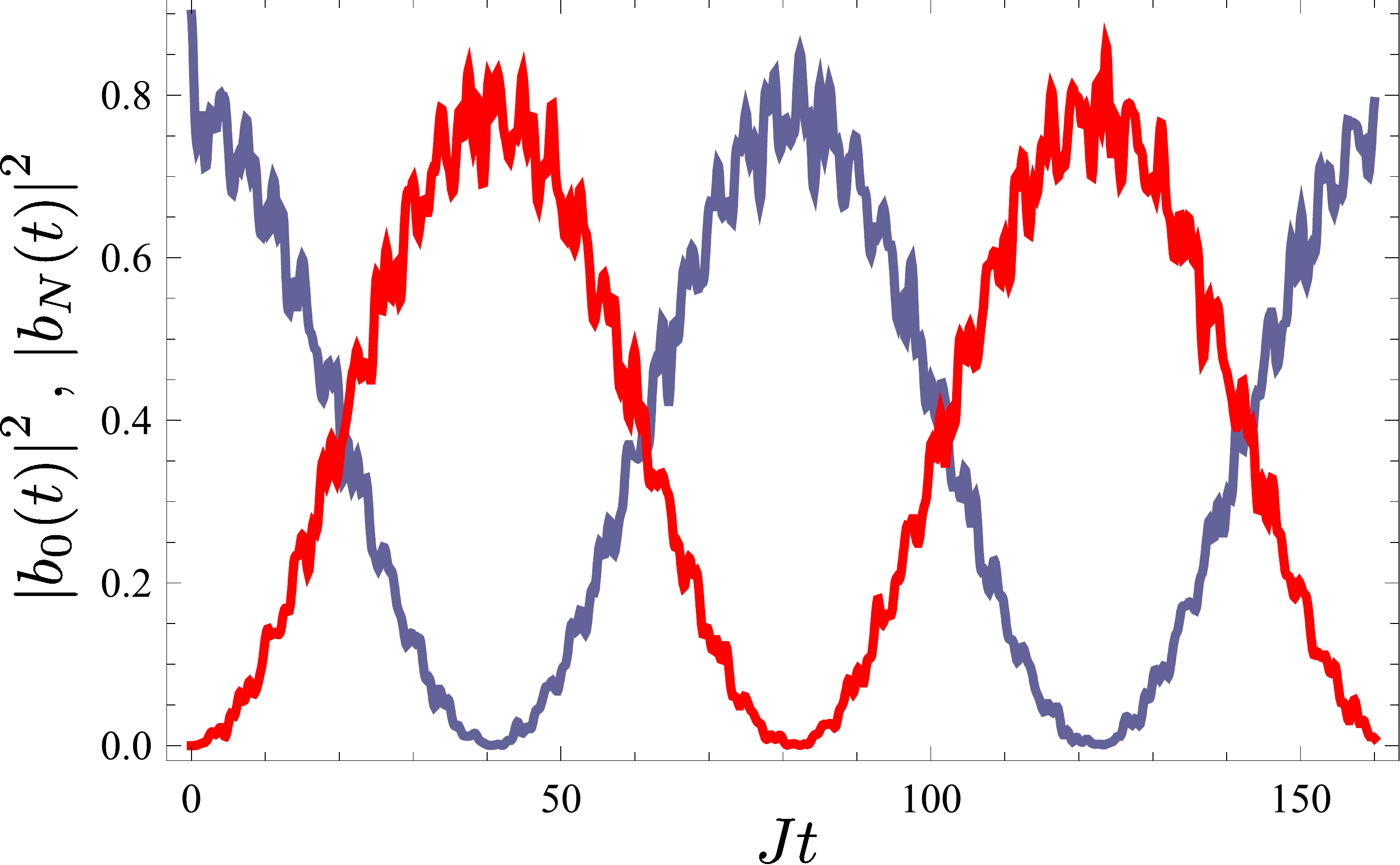}
 \hfill
 \textbf{(b)} \includegraphics[width=0.45\columnwidth]{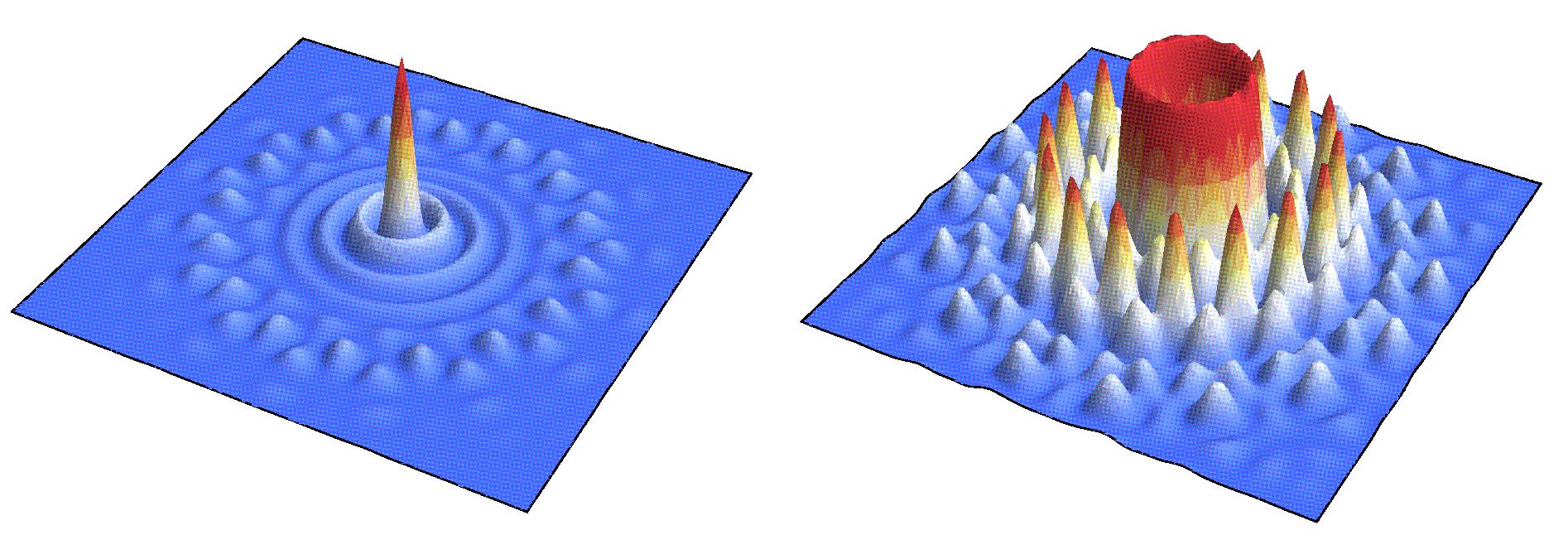}
 \caption{
\textbf{(a)} Many-body dynamics $|b_0(t)|^2$ and $|b_N(t)|^2$ for $L=4$ and $N=8$ following a sudden change in phase twist from $\theta = 0$ to $\theta=\pi/4$ at $U/J =1$ and $t/J = 0.7$.
\textbf{(b)} Transverse time-of-flight images of an initial state with quasi-momentum $q_0=0$ and $q_0=L/4$ for $L=16$.}
\label{fig:dynamics}
\end{figure}

In Fig.~\ref{fig:dynamics}(a) we show the many-body dynamics, i.e.~$|b_0(t)|^2$ and $|b_N(t)|^2$ for $L=4$ and $N=8$ following a sudden change in phase twist from $\theta = 0$ to $\theta=\pi/4$ at time $t=0$ with $U/J =1$ and $t/J = 0.7$. We see that the many-body oscillations following this sudden change occur at the frequency of the gap $\hbar \nu$. These oscillations are modulated by an oscillation with frequency $U (N-1)/L$ which is the energy difference between the nearly-degenerate ground states and the next lowest-lying intermediate states $\ket{2,N-2}$ and $\ket{N-2,2}$. The amplitude of the oscillations is smaller than one, i.e. $|b_0(t)|^2 + |b_N(t)|^2 < 1$, due to finite depletion of the cat state.

In Fig.~\ref{fig:dynamics}(b) we show the numerically calculated time-of-flight images for two states with initial quasi-momentum $q_0 = 0$ and $L/4$, respectively. These states correspond to the two types of initial distributions that one has to experimentally distinguish in order to probe the many-body oscillations. The figure shows the distinctive interference pattern of the two initial states.

In the limit of $L \rightarrow \infty$ lattice sites the time-of-flight profile reduces to
\begin{eqnarray}
|\psi( \mathbf{r}, t \to \infty )|^2 \propto \left|J_{q_0}\left( \frac{M \mathbf{r}} {\hbar t R} \right) \right|^2
\end{eqnarray}
where $J_{q_0}(x)$ are Bessel functions of the first kind and $R$ denotes the radius of the ring lattice. This case corresponds to the rotationally symmetric case discussed in Ref.~\cite{Cozzini06} in which a state initially with zero quasi-momentum for example will exhibit an interference peak at the origin while a state with non-zero quasi-momentum will exhibit a central hole. For a finite number of lattice sites the sum does not correspond exactly to a Bessel function and the momentum distribution does not become fully radially symmetric. However, there is still a unique correspondence between the position of the peaks in the absorption image and the initial quasi-momentum distribution (see also Ref.~\cite{Peden07}). Consequently, the latter provides a means to experimentally determine the quasi-momentum of the wavefunction before the release.

\section{Conclusion}

We have investigated the effects of rotation on the ground state properties of bosonic atoms in a rotating ring lattice. In the uniform geometry only the commensurate (CO) system exhibites a cat state at the critical phase twist and its gap scales with the number of particles $N$ as $(U/2L)^{\bar{n}(L-1)}$. Introducing a superlattice lifts the CO constraint and improves the scaling to $U (\eta/2)^{N/2}$. This makes the cat less sensitive to imperfections in the system. Finally, we have proposed to probe the cat-like correlations via the many-body oscillations induced by a sudden change in the rotation frequency. Since the exponential scaling of the energy gap with the number of particles remains, we suspect that cat state production in ultracold atomic gases is limited to modest number of atoms at least for systems with contact interactions. One possible way-out is to consider systems with long-range interactions such as polar molecules with dipolar interactions.

\section{Acknowledgments}

We are grateful to David Hallwood for useful discussions. This work was partially supported by the National Science Foundation through a grant for the Institute for Theoretical Atomic, Molecular and Optical Physics (ITAMP) at Harvard University and Smithsonian Astrophysical Observatory. A.N.~is grateful for the hospitality of ITAMP visitor's program and acknowledges support from the Rhodes Trust.

\end{document}